\documentclass[12pt]{article}
\begin{document}
\renewcommand{\thesection}{\Roman{section}}
\baselineskip=22 pt
\setcounter{page}{1}
\thispagestyle{empty}
\topskip  -2.5  cm
\vspace{0.2 cm}

\centerline{\Large \bf Speed of Sound in the Mass Varying Neutrinos
Scenario}
\vskip 0.8 cm
\centerline{{\large Ryo Takahashi$^{a}$}$\!\!\!$
\renewcommand{\thefootnote}{\fnsymbol{footnote}}
\footnote[2]{e-mail:  takahasi@muse.sc.niigata-u.ac.jp} \
{\large and Morimitsu Tanimoto$\,^{b}$}$\!\!\!$
\renewcommand{\thefootnote}{\fnsymbol{footnote}}
\footnote[3]{e-mail: tanimoto@muse.sc.niigata-u.ac.jp}}

\vskip 0.5 cm

\centerline{$^a\!$ Graduate School of  Science and Technology,
 Niigata University,  950-2181 Niigata, Japan}

\centerline{$^b\!$ Department of Physics,
Niigata University,  950-2181 Niigata, Japan}

\vskip 2 cm
\centerline{\bf ABSTRACT}\par
\vskip 0.2 cm

We discuss about the speed of sound squared in the Mass Varying
Neutrinos scenario (MaVaNs). Recently, it was argued that the MaVaNs
has a catastrophic instability which is the emergence of an imaginary
speed of sound at the non-relativistic limit of neutrinos. As the
result of this instability, the neutrino-acceleron fluid cannot act as
the dark energy. However, it is found that the speed of sound squared
in the neutrino-acceleron fluid could be positive in our model. We
examine the speed of sound in two cases of the scalar potential. One
is the small fractional power-law potential and another is the
logarithmic one. The power-law potential model with the right-handed
neutrinos gives a stable one.

\newpage

\section{INTRODUCTION}
One of the most challenging questions in both cosmological and 
particle physics is the nature of the dark energy in the Universe.
At the present epoch, the energy density of the Universe is dominated 
by a dark energy component, whose negative pressure causes the
expansion of the Universe to accelerate. In order to clarify the
origin of the dark energy, one has tried to understand the connection
of the dark energy with particle physics.

In a scenario proposed by Fardon, Nelson and Winer (MaVaNs), relic
neutrinos could form a negative pressure fluid and cause cosmic
acceleration \cite{Weiner}. In this idea, an unknown scalar field which
is called ``acceleron'' is introduced and neutrinos interact through a
new scalar force. The acceleron field sits at the instantaneous
minimum of its potential, and the cosmic expansion only modulates this
minimum through changes in the neutrino density. Therefore the
neutrino mass is given by the acceleron, in other words, it depends on
its number density and changes with the evolution of the Universe. The
cosmological parameter $w$ and the dark energy also evolve with the
neutrino mass. Those evolutions depend on a model of the scalar
potential strongly. Typical examples of the potential have been
discussed by Peccei \cite{Peccei}. 

The variable  neutrino mass  was considered at first in 
\cite{Yanagida}, and was discussed for neutrino clouds
\cite{McKellar}. Ref. \cite{Wang} considered coupling of the dark
energy scalar, such as Quintessence to the neutrinos and discuss its
impact on the neutrino mass limits from Baryogenesis.
The MaVaNs scenario leads to interesting  phenomenological results.
 The neutrino oscillations may be a probe of the dark energy
 \cite{Kaplan,Mini,Feng}. The leptogenesis \cite{Bi,Gu}, the cosmo MSW
 effect of neutrinos \cite{Hung} and the solar neutrino
 \cite{SUN1,SUN2} have been studied in the context of this scenario.
 Cosmological  discussions of  the  scenario are also presented
 \cite{Zhang,Bruck,Right,WZ,Supernova,Mota,11027}. The extension to
the supersymmetry have been presented in ref. \cite{our,SUSY}. This
scenario is also discussed in the context of the texture of the
neutrino mass matrix with three families
 \cite{Honda}.

Despite many implications of this scenario, ref. \cite{stability}
showed that this scenario contains a catastrophic instability which
occurs when neutrinos become non-relativistic. As neutrinos become
non-relativistic, the speed of sound squared in the neutrino-acceleron 
fluid turns to be negative. As the result of this instability,
neutrinos condense into neutrino nuggets, and thus cannot act as the
dark energy. Here, it is important to say that some non-adiabatic
models \cite{Zhang,Bruck} do not suffer from this
instability. However, we have found
that the speed of sound squared in this fluid could be positive even
though  neutrinos are not enough relativistic and the
neutrino-acceleron fluid is adiabatic. In order to realize the
positive speed of sound squared, a constraint for the scalar potential
is required.

The paper is organized as follows: in Sec.II, we summarize the MaVaNs
scenario with three families. Sec.III presents discussions for the
speed of sound in the hydrodynamic picture. Sec.IV presents a stable
model in the MaVaNs scenario. Sec.V devotes to the summary.
 
\section{DARK ENERGY FROM MaVaNs}
In the MaVaNs scenario, one considers a dark energy sector consisting
an ``acceleron'' field, $\phi_a$ and a dark fermion, $\psi _n$. This
sector couples to the standard model sector only through
neutrinos. The dark energy is assumed to be the sum of the energy
densities of neutrinos and a scalar potential for the  acceleron:
 \begin{equation}
  \rho _{\mbox{{\scriptsize DE}}}=\rho _\nu +V(\phi _a),
  \label{DE}
 \end{equation} 
where the potential energy of the acceleron is responsible for the
acceleration of the Universe and for the dynamical neutrino mass. The
energy density for three generations of neutrinos and antineutrinos is
generally given by
 \begin{equation}
  \rho _\nu =T^4\sum_{i=1}^3F(\xi _i),\hspace{0.5cm}
  \xi _i\equiv\frac{m_{\nu i}}{T},\hspace{0.5cm}
  F(\xi _i)\equiv\frac{1}{\pi ^2}
                 \int_0^\infty\frac{dyy^2\sqrt{y^2+\xi _i^2}}{e^y+1},
  \label{rhonu}
 \end{equation}
where $i$ denotes three families.\footnote{In
ref. \cite{Weiner,Peccei}, there is a more precise discussion about
the Fermi factor in eq.(\ref{rhonu}). For our purpose it suffices to
use this form.}

In the scenario, $\rho _{\mbox{{\scriptsize DE}}}$ is stationary with
respect to variation in the neutrino mass. This stationary condition
is represented by
 \begin{equation}
  \frac{\partial\rho _\nu}
       {\partial\displaystyle\sum_{i=1}^3m_{\nu i}}
  +\frac{\partial V(\phi _a(m_{\nu i}))}
        {\partial\displaystyle\sum_{i=1}^3m_{\nu i}}=0.
  \label{stationary}
 \end{equation}
If $\partial\sum m_{\nu i}/\partial\phi _a\neq 0$, this condition
turns to
 \begin{equation}
  T^4\displaystyle\sum_{i=1}^3
  \frac{\partial F}{\partial\xi _i}
  \frac{\partial\xi _i}{\partial\phi _a}
  +\frac{\partial V(\phi _a)}{\partial\phi _a}=0.
  \label{stationary1}
 \end{equation}
Using the equation of energy conservation in the Robertson-Walker
background and the above stationary condition, one can get the
equation of state parameter $w$ as follows:
 \begin{equation}
  w+1=\frac{\left[4-h(T)\right]\rho _\nu}
           {3\rho _{\mbox{{\scriptsize DE}}}},
  \label{w}
 \end{equation}
where 
 \begin{equation}
  h(T)\equiv\frac{\displaystyle\sum_{i=1}^3\xi _i
            \frac{\partial F(\xi _i)}{\partial\xi _i}}
            {\displaystyle\sum_{j=1}^3F(\xi _j)}.
  \label{h}
 \end{equation}

The speed of sound squared in the neutrino-acceleron fluid is given by
 \begin{equation} 
  c_s^2=\frac{\dot{p}}
             {\dot{\rho}_{\mbox{{\scriptsize DE}}}}
       =\frac{\dot{w}\rho _{\mbox{{\scriptsize DE}}}
        +w\dot{\rho}_{\mbox{{\scriptsize DE}}}}
         {\dot{\rho}_{\mbox{{\scriptsize DE}}}}
 \label{cs}
 \end{equation}
where $p$ is the pressure of the dark energy
\cite{stability,Mukhanov}. Recently, it was argued that when neutrinos
are non-relativistic, this speed of sound squared becomes negative in
this scenario: $c_s^2=(\partial\ln m_\nu /\partial\ln n_\nu )<0$,
where $n_\nu$ is the number density of neutrinos. The emergence of an
imaginary speed of sound shows that the MaVaNs scenario with
non-relativistic neutrinos is unstable, and thus the fluid in the
scenario cannot act as the dark energy. However, it is found that the
speed of sound squared in this fluid can be positive even though
 neutrinos are non-relativistic. Then, a constraint for the scalar
 potential is required.

\section{SPEED OF SOUND}
At the non-relativistic limit of neutrinos, the energy density of
neutrinos is given by
 \begin{equation}
  \rho _\nu =\displaystyle\sum_{i=1}^3m_{\nu i}n_\nu,
  \label{rhononnu}
 \end{equation}
then the stationary condition eq.(\ref{stationary}) is rewritten as
follows:
 \begin{equation}
  n_\nu=-\frac{\partial V(\phi _a)}
              {\partial\displaystyle\sum_{i=1}^3m_{\nu i}}.
 \label{stationary2}
 \end{equation}
Now the dark energy density is given as
 \begin{equation}
  \rho _{\mbox{{\scriptsize DE}}}
  =\displaystyle\sum_{i=1}^3m_{\nu i}n_\nu+V(\phi _a).
  \label{DE1}
 \end{equation}
Using these relations, the speed of sound squared in the
neutrino-acceleron fluid is 
negative: $c_s^2=(\partial\ln m_\nu /\partial\ln n_\nu )<0$ as shown
in ref. \cite{stability}. In order to study $c_s^2$ quantitatively, we
start with discussing the energy density of neutrinos in
eq.(\ref{rhonu}). Taking account that $\xi _i$ is much larger than $1$
for non-relativistic neutrinos, the function $F(\xi _i)$ is expanded
in terms of $\xi _i^{-1}$ as: 
 \begin{equation}
  F(\xi _i)\simeq\frac{\hat{n}_\nu }{T^3}\xi _i
                 +a\frac{\hat{n}_\nu }{T^3}\frac{1}{\xi _i},
  \label{F1}
 \end{equation}
where
 \begin{equation}
  \hat{n}_\nu 
  \equiv\frac{T^3}{\pi ^2}\int_0^\infty\frac{dyy^2}{e^y+1},
  \hspace{1cm}a\equiv\frac{\int_0^\infty\frac{dyy^4}{e^y+1}}
  {2\int_0^\infty\frac{dyy^2}{e^y+1}}\simeq 6.47. 
  \label{n0}
 \end{equation}
Since the first term of the right hand side in eq.(\ref{DE1}) is
derived from the first term of the right hand side in eq.(\ref{F1}),
the effect of the second term in eq.(\ref{F1}) should be added to the
dark energy density as a correction. Due to this correction, it could
be that the negative speed of sound squared turns to be positive under
a condition, which is discussed in detail later. 

The dark energy density including the correction term is given as
follows:
 \begin{equation}
  \rho _{\mbox{{\scriptsize DE}}}
  =\displaystyle\sum_{i=1}^3m_{\nu i}\hat{n}_\nu
   \left(1+\frac{a}{\xi _i^2}\right)+V(\phi _a).
  \label{DE2}
 \end{equation}
Then the stationary condition eq.(\ref{stationary1}) is described as
 \begin{equation}
  T\displaystyle\sum_{i=1}^3
  \left(\hat{n}_\nu-\frac{a\hat{n}_\nu}{\xi _i^2}\right)
  \frac{\partial\xi _i}{\partial\phi _a}
  +\frac{\partial V(\phi _a)}{\partial\phi _a}=0.
  \label{stationary3}
 \end{equation}
The equation of state is generally given by eq.(\ref{w}). At the
 non-relativistic limit, it is easy to see $h(T)=1$. Because of the
correction term, $h(T)$ is deviated from $1$ as follows:
 \begin{equation}
  h(T)\equiv\frac{T^4\displaystyle\sum_{i=1}^3\xi _i
            \frac{\partial F(\xi _i)}{\partial\xi _i}}
            {\displaystyle\sum_{j=1}^3F(\xi  _j)}
      =\frac{1-\displaystyle\frac{1}
                    {\displaystyle\sum_{i=1}^3\xi _i}
       \left(\displaystyle\sum_{j=1}^3\frac{a}{\xi _j}\right)}
       {1-\displaystyle\frac{1}{\displaystyle\sum_{k=1}^3\xi _k}
       \left(\displaystyle\sum_{l=1}^3\frac{a}{\xi _l}\right)}
      \simeq 1-\frac{2}{\displaystyle\sum_{i=1}^3\xi _i}
             \left(\displaystyle\sum_{j=1}^3\frac{a}{\xi _j}\right).
  \label{h1}
 \end{equation}
Thus, the equation of state is given by
 \begin{equation}
  w+1=\frac{\hat{n}_\nu\displaystyle\sum_{i=1}^3\left(
      3m_{\nu i}+\frac{5aT}{\xi _i}\right)}
      {3\rho _{\mbox{{\scriptsize DE}}}}
  \label{w1} 
 \end{equation}
where we omitted the term of $\mathcal{O}(1/\xi ^3)$. Using
eqs.(\ref{DE2}) and (\ref{w1}) and the stationary condition
 eq.(\ref{stationary}), the speed of sound squared in the
neutrino-acceleron fluid is described finally as follows (see
Appendix):
 \begin{eqnarray}
  c_s^2&=&\frac{\dot{w}\rho _{\mbox{{\scriptsize DE}}}
        +w\dot{\rho}_{\mbox{{\scriptsize DE}}}}
         {\dot{\rho}_{\mbox{{\scriptsize DE}}}}
       =\frac{\frac{\partial w}{\partial z}
        \rho _{\mbox{{\scriptsize DE}}}
        +w\frac{\partial \rho _{\mbox{{\scriptsize DE}}}}{\partial z}}
        {\frac{\partial \rho _{\mbox{{\scriptsize DE}}}}{\partial z}}
        \nonumber\\
       &=&\frac{\displaystyle\sum_{i=1}^3
        \frac{\partial m_{\nu i}}{\partial z}\hat{n}_\nu}
        {\displaystyle\sum_{i=1}^3m_{\nu i}
        \frac{\partial \hat{n}_\nu }{\partial z}}
        +\frac{\frac{5}{3}a\hat{n}_\nu\displaystyle\sum_{i=1}^3
         \left(\frac{5T_0}{\xi _i}
        -\frac{T}{\xi _i^2}
        \frac{\partial m_{\nu i}}{\partial z}\right)}
        {\displaystyle\sum_{i=1}^3m_{\nu i}
        \frac{\partial\hat{n}_\nu }{\partial z}},
  \label{cs1}
 \end{eqnarray}
where $z$ is the redshift parameter, $z\equiv (T/T_0)-1$ and ``$0$''
represents a value at the present epoch. We have taken the form
differentiated by the redshift parameter instead of the time. The
first term in the right hand side of eq.(\ref{cs1}) is the leading
term at the non-relativistic limit and is negative definite because of
$\partial m_{\nu i}/\partial z<0$ and $\partial\hat{n}_\nu /\partial
z>0$. The numerator of the second term, which is the correction term,
is the positive definite. Thus, it is possible that the speed of sound
squared is positive, in other words, the neutrino-acceleron fluid
could be stable due to this term. From the eq.(\ref{cs1}), it is easy
to see that if the following relation is satisfied, the speed of sound
squared becomes positive:
 \begin{equation}
  \displaystyle\sum_{i=1}^3
  \frac{\partial m_{\nu i}}{\partial z}
  \left(1-\frac{5aT^2}{3m_{\nu i}^2}\right)
  +\frac{25aT_0^2(z+1)}{3}\displaystyle\sum_{i=1}^3
  \frac{1}{m_{\nu i}}>0.
  \label{constraint}
 \end{equation}
The sign of $c_s^2$ is determined by the magnitude of $\partial m_{\nu
i}/\partial z$. Since the magnitude of $\partial m_{\nu i}/\partial z$
is model dependent, we will examine two cases of the scalar potential
for the acceleron, one is the power-law potential and another is the
logarithmic one in the next section. 

\section{MODELS}

In the MaVaNs scenario, one needs a flat scalar potential
\cite{Weiner}. Therefore, we will discuss the speed of sound in
cases of the small fractional power-law potential and logarithmic one
in this section. The magnitude of $\partial m_{\nu i}/\partial z$
does not only depend on the scalar potential but also the coupling
between neutrinos and the acceleron. Therefore, we consider two cases
in eqs.(\ref{3l1n}) and (\ref{3l3r1n}) for this coupling.

\subsection*{A. The power-law potential}
We take the small fractional power-law potential of the form
 \begin{eqnarray}
  V(\phi _a)
   =A\left(\frac{\phi _a}{\phi _a^0}\right)^k,\hspace{5mm}
    k\ll 1,\label{power}
 \end{eqnarray}
where parameters $A$ and $k$ are fixed by the magnitude of the dark
 energy and the stationary condition at the present epoch,
 respectively.

\subsubsection*{Three left-handed neutrinos and a sterile neutrino}

We take a Lagrangian of the form
 \begin{equation}
  \mathcal{L}=\bar{\nu}_{L\alpha}m^\alpha _D\psi _n+\lambda\phi _a\psi 
              _n\psi _n+h.c.~,
  \label{3l1n}
 \end{equation}
where $\nu _L$ and $\psi _n$ are the left-handed and a sterile
 neutrino, respectively. Since we consider three families of active
 neutrinos, the mass matrix $m_D$ is $3\times 1$ matrix. Thus the
 neutrino mass matrix is given by
 \begin{equation}
  M_\nu=\frac{m_Dm_D^T}{\lambda\phi _a},
  \label{mm}
 \end{equation}
where we assume $m_D^\alpha\ll\lambda\phi _a$. After diagonarizing
 this matrix, we can find mass eigenvalues of neutrinos as follows:
 \begin{equation}
  m_{\nu i}=\frac{M_i^2}{\lambda\phi _a},
  \label{numass}
 \end{equation}
which gives
 \begin{equation}
  \xi _i=\frac{M_i^2}{\lambda\phi _aT}.\label{xi}
 \end{equation}
Using eqs.(\ref{DE}), (\ref{power}) and (\ref{xi}), we have the
 redshift dependence of the acceleron from the stationary condition
 eq.(\ref{stationary1}) as
 \begin{equation}
  \phi _a\simeq
  \frac{-Ak+G(z)}
  {2\lambda a\hat{n}_\nu T^2\displaystyle\sum_{l=1}^3(1/M_l^2)},
  \hspace{5mm}G(z)\equiv\sqrt{A^2k^2+4a\hat{n}_\nu ^2T^2
  \displaystyle\sum_{j=1}^3(1/M_j^2)
  \displaystyle\sum_{k=1}^3M_k^2},
  \label{phia1}
 \end{equation}
where the leading term is taken after expanding $\partial
 V/\partial\phi _a$ by $k$. Thus, the redshift
dependence of neutrino masses is given by\newpage
 \begin{eqnarray}
  \displaystyle\sum_{i=1}^3
  \frac{\partial m_{\nu i}}{\partial z}
  &=&-\displaystyle\sum_{i=1}^3\frac{M_i^2}{\lambda\phi _a^2}
     \frac{\partial\phi _a}{\partial z}\nonumber\\
  &=&-\frac{4a^2\hat{n}_\nu ^2T^4
     \left[\displaystyle\sum_{l=1}^3(1/M_l^2)\right]^2
     \displaystyle\sum_{i=1}^3M_i^2}
     {(z+1)[-Ak+G(z)]^2}
     \left[\frac{8\hat{n}_\nu\displaystyle\sum_{l=1}^3M_l^2}
     {G(z)}
     -\frac{5[-Ak+G(z)]}
     {2a\hat{n}_\nu T^2\displaystyle\sum_{l=1}^3
     (1/M_l^2)}\right],
  \label{mz3}
 \end{eqnarray}
where the right hand side is negative and its absolute value is
 the increasing function of $k$.

At the present epoch, values of some parameters are given by
 \begin{equation}
  T_0\simeq 1.69\times 10^{-4}(\mbox{eV}),\hspace{5mm} 
  A\simeq 2.99\times 10^{-11}(\mbox{eV}^4),\hspace{5mm}
  \hat{n}_\nu ^0\simeq 8.82\times 10^{-13}(\mbox{eV}^3),
  \label{values1}
 \end{equation}
and we take the following typical masses at the present epoch: 
 \begin{equation}
  m_{\nu 1}^0=0.0045(\mbox{eV}),\hspace{5mm}
  m_{\nu 2}^0=0.01(\mbox{eV}),\hspace{5mm}
  m_{\nu 3}^0=0.05(\mbox{eV}),
  \label{masses1}
 \end{equation}
which lead to $\Delta m_{\mbox{atm}}^2=2.4\times 10^{-3}(\mbox{eV}^2)$
and $\Delta m_{\mbox{sun}}^2=8.0\times 10^{-5}(\mbox{eV}^2)$. We
 numerically evaluate the relation (\ref{constraint}) by putting
 values of (\ref{values1}) and (\ref{masses1}). We find that $k$ has
 to be smaller than $5.5\times 10^{-5}$ to satisfy the relation
 (\ref{constraint}). However this value of $k$ is unfavored in the
phenomenology of the neutrino experiments. The value of $k$ is related 
with neutrino masses through the stationary condition
eq.(\ref{stationary3}). Actually, the $k$ which is smaller than
$5.5\times 10^{-5}$ leads to $\sum m_{\nu
i}\sim\mathcal{O}(10^{-4})$(eV). In the case of $10^{-4}<k<10^{-2}$,
$\sum m_{\nu i}\sim\mathcal{O}(10^{-3}\sim 10^{-1})$(eV) is
expected. Therefore, this model including the relation (\ref{numass})
and the scalar potential (\ref{power}) is unfavored.

\subsubsection*{Three left- and right-handed neutrinos and a sterile
 neutrino}
We add usual right-handed neutrinos to the Lagrangian (\ref{3l1n})
 \begin{equation}
  \mathcal{L}=\bar{\nu}_{L\alpha}m^\alpha _D\psi _n
              +\lambda\phi _a\psi _n\psi _n
              +\bar{\nu}_{L\alpha}M^{\alpha\beta}_D\nu _{R\beta}
              +\nu ^T_{R\alpha}M^{\alpha\beta}_RC^{-1}\nu _{R\beta}
              +h.c.~,
  \label{3l3r1n}
 \end{equation}
where $\nu _R$ is the right-handed neutrinos and both $M_D$ and $M_R$
are $3\times 3$ matrix. Then, the neutrino mass matrix is given as the
$7\times 7$ matrix
 \begin{equation}
  M=\left(
   \begin{array}{ccc}
    0     & m_D            & M_D \\
    m^T_D & \lambda\phi _a & 0 \\
    M^T_D & 0              & M_R
   \end{array}
  \right),
  \label{mm1}
 \end{equation}
in the $(\nu _L,\psi _n, \nu _R)$ basis. We take the right-handed
Majorana mass scale to be much higher than the Dirac neutrino mass
scale, and assume $m_D^\alpha\ll\lambda\phi _a$. Then, the effective
neutrino mass matrix is approximately given by
 \begin{equation}
  M_\nu=M_DM^{-1}_RM^T_D+\frac{m_Dm_D^T}{\lambda\phi _a}.
  \label{mm2}
 \end{equation}

The first term in the right hand side of eq.(\ref{mm2}) is the
time-independent neutrino seesaw mass matrix, which is denoted by
$\tilde{M}_\nu$, and so it depends on the flavor model of
neutrinos. In this case, since the neutrino mass matrix reduces to
eq.(\ref{mm}) at the limit of $\tilde{M}_\nu\rightarrow 0$, we assume
that $\tilde{M}_\nu$ dominates the effective neutrino mass matrix
$M _\nu$. Then we can describe generally mass eigenvalues in the first 
order perturbation as follows \cite{Honda}:
 \begin{equation}
  m_{\nu i}=\tilde{m}_{\nu i}+c_i\frac{M_i^2}{\lambda\phi _a},
  \label{numass1}
 \end{equation}
where $c_i$ is a coefficient of order $1$ depending on the model of
families.\footnote{The values of $c_i$ are given in a specific flavor
symmetry as in ref. \cite{Honda}.} Using this relation and
eq.(\ref{power}), we have the redshift dependence of the acceleron
from the stationary condition eq.(\ref{stationary3}) as
\begin{equation}
  \phi _a\simeq
  \frac{\hat{n}_\nu\displaystyle\sum_{i=1}^3c_iM_i^2
  \left(1-\frac{aT^2}{\tilde{m}_{\nu i}^2}\right)}
  {\lambda kA},
 \end{equation}
where the first term of the left hand side in eq.(\ref{stationary3})
was expanded by the second term of the right hand side in
eq.(\ref{numass1}). Thus, the redshift dependence of neutrino masses
  is given by
 \begin{eqnarray}
  \displaystyle\sum_{i=1}^3
  \frac{\partial m_{\nu i}}{\partial z}
  =-\frac{kA\displaystyle\sum_{i=1}^3c_iM_i^2
            \displaystyle\sum_{j=1}^3M_j^2
            \left(3-\frac{5aT^2}{\tilde{m}_{\nu j}^2}\right)}
      {\hat{n}_\nu (z+1)\left[\displaystyle\sum_{k=1}^3M_k^2
      \left(1-\frac{aT^2}{\tilde{m}_{\nu k}^2}\right)\right]^2}.
 \end{eqnarray}
The magnitude of the first term of the left hand side in
  eq.(\ref{constraint}) is nearly equal to $kA/\hat{n}_\nu$ at the
  present epoch. In order to realize the positive speed of sound
  squared, $k$ has to be smaller than $5.19\times 10^{-6}$. When we
  assume that the magnitude of the second term of the right hand side
  in eq.(\ref{numass1}) is $0.1$ percent of the first term, we obtain
  $k=1.90\times 10^{-6}$ which reproduce observed values of neutrino
  masses. Therefore, the model including the scalar potential
  (\ref{power}) and the relation between neutrino masses and the
  acceleron (\ref{numass1}) is favored in the MaVaNs scenario and can
  act as the dark energy.  

\subsection*{B. The logarithmic potential}
We take the logarithmic scalar potential of the form
 \begin{eqnarray}
  &&V(\phi _a)=B\ln 
             \left(\frac{\phi _a}{\mu}\right),\label{log}\\
  &&B\simeq 5.68\times 10^{-14}(\mbox{eV}^4),\label{B}
 \end{eqnarray}
where we use values of (\ref{values1}) and (\ref{masses1}) to fix the
 value of $B$.

We will consider two cases of the coupling between neutrinos and 
the acceleron as well as the case of the small fractional potential.

\subsubsection*{Three left-handed neutrinos and a sterile
 neutrino}

We take Lagrangian in eq.(\ref{3l1n}), and thus, neutrino
masses in eq.(\ref{numass}). The stationary condition of
eq.(\ref{stationary3}) gives the redshift dependence of the acceleron:
 \begin{equation}
 \phi _a\simeq
  \frac{-B+H(z)}
  {2\lambda a\hat{n}_\nu T^2\displaystyle\sum_{l=1}^3(1/M_l^2)},
  \hspace{5mm}H(z)\equiv\sqrt{B^2+4a\hat{n}_\nu ^2T^2
  \displaystyle\sum_{j=1}^3(1/M_j^2)
  \displaystyle\sum_{k=1}^3M_k^2},
  \label{phia2}
 \end{equation}
and thus the redshift dependence of neutrino masses is given by
\begin{eqnarray}
  \displaystyle\sum_{i=1}^3
  \frac{\partial m_{\nu i}}{\partial z}
  =-\frac{4a^2\hat{n}_\nu ^2T^4
   \left[\displaystyle\sum_{l=1}^3(1/M_l^2)\right]^2
   \displaystyle\sum_{i=1}^3M_i^2}{(z+1)[-B+H(z)]^2}
   \left[\frac{8\hat{n}_\nu\displaystyle\sum_{l=1}^3M_l^2}
   {H(z)}
   -\frac{5\{-B+H(z)\}}
   {2a\hat{n}_\nu T^2\displaystyle\sum_{l=1}^3
   (1/M_l^2)}\right].
  \label{mz2}
 \end{eqnarray}
Using values of eqs.(\ref{values1}), (\ref{masses1}) and (\ref{B}),
  the first term of the left hand side in eq.(\ref{constraint}) is
  $-\mathcal{O}(10^{-1})$ $(\mbox{eV})$, however the second term is
  $\mathcal{O}(10^{-4})(\mbox{eV})$ at the present epoch. Thus the
  speed of sound squared becomes negative, in other words, one cannot
  build a stable MaVaNs model including the form of the neutrino mass
  like as eq.(\ref{numass}).

\subsubsection*{Three left- and right-handed neutrinos and a sterile
 neutrino}
We take Lagrangian in eq.(\ref{3l3r1n}), and thus, neutrino
masses in eq.(\ref{numass1}). The redshift dependence of the acceleron 
is given as:
\begin{equation}
  \phi _a\simeq
  \frac{\hat{n}_\nu\displaystyle\sum_{i=1}^3c_iM_i^2
  \left(1-\frac{aT^2}{\tilde{m}_{\nu i}^2}\right)}
  {\lambda B},
  \label{phia} 
 \end{equation}
and thus, the redshift dependence of neutrino masses
is given by
 \begin{eqnarray}
  \displaystyle\sum_{i=1}^3
  \frac{\partial m_{\nu i}}{\partial z}
  =-\frac{B\displaystyle\sum_{i=1}^3c_iM_i^2
            \displaystyle\sum_{j=1}^3M_j^2
            \left(3-\frac{5aT^2}{\tilde{m}_{\nu j}^2}\right)}
      {\hat{n}_\nu (z+1)\left[\displaystyle\sum_{k=1}^3M_k^2
      \left(1-\frac{aT^2}{\tilde{m}_{\nu k}^2}\right)\right]^2}.
  \label{mz}
 \end{eqnarray}
Using values of eqs.(\ref{values1}) and (\ref{masses1}), we can
estimate the redshift dependence of neutrino masses. Since
$\tilde{m}_{\nu i}$ dominate the neutrino mass, we assume
$\tilde{m}_{\nu i}\sim m_{\nu i}$. In this case, the first term of the
left hand side in eq.(\ref{constraint}) is
$-\mathcal{O}(10^{-3})$(eV), however the second term is
$\mathcal{O}(10^{-4})$(eV). Thus, the speed of sound squared becomes
negative as well as the previous case. We conclude that the speed of
sound squared in the neutrino-acceleron fluid becomes negative when
neutrinos are non-relativistic for the logarithmic scalar potential in
  eq.(\ref{log}) and neutrino masses like as eqs.(\ref{numass}) and
  (\ref{numass1}).

\section{SUMMARY}
We have discussed about the speed of sound squared in the
neutrino-acceleron fluid, and tried to find a stable MaVaNs model in
which the fluid is adiabatic. In order to examine $c_s^2$
quantitatively, we have taken two types of the scalar potential for
the acceleron. One is the small fractional power-law potential and
another is the logarithmic one. Furthermore, we have studied about two
types of models which have different couplings between neutrinos and
the accerelon. 

In our analysis, models including the logarithmic scalar potential are
unstable and cannot act as the dark energy because the speed of sound
squared in the neutrino-acceleron fluid becomes negative. However,
models with the small fractional power-law potential can avoid this
instability. The model including only three left-handed neutrinos and
a sterile neutrino avoids this instability but does not reproduce the
observed neutrino masses. On the other hand, the model including the
right-handed neutrinos reproduces the observed neutrino masses and
realizes the positive speed of sound squared. Neutrino masses in this
model have the time-independent component from the seesaw mechanism,
which was assumed to be dominant in the effective neutrino
mass. Therefore, it is easy to reconcile these neutrino masses with
observed ones. Due to this time-independent mass, this model becomes
viable.

\section*{Appendix}
The derivation of the speed of sound squared
in eq.(\ref{cs1}) is presented in this Appendix.

The energy density for three generations of neutrinos and
antineutrinos is generally given by
 \begin{equation}
  \rho _\nu =T^4\sum_{i=1}^3F(\xi _i),\hspace{5mm}
  \xi _i\equiv\frac{m_{\nu i}}{T},\hspace{5mm}
  F(\xi _i)\equiv\frac{1}{\pi ^2}
                 \int_0^\infty\frac{dyy^2\sqrt{y^2+\xi _i^2}}{e^y+1}.
 \label{rhonu1}
 \end{equation}
As neutrinos become non-relativistic, $\xi _i$ is much larger than
$1$. Therefore, the function $F(\xi _i)$ is expanded in terms of $\xi
_i^{-1}$ as: 
 \begin{eqnarray}
  F(\xi _i)&=&\frac{1}{\pi ^2}
              \int_0^\infty\frac{dyy^2}{e^y+1}
              \xi _i\sqrt{\left(\frac{y}{\xi i}\right)^2+1}\nonumber\\
           &\simeq&\frac{1}{\pi ^2}
                   \int_0^\infty\frac{dyy^2}{e^y+1}
                   \xi _i\left[\frac{1}{2}\left(
                   \frac{y}{\xi _i}\right)^2+1\right]\nonumber\\
           &=&\frac{\xi _i}{\pi ^2}\int_0^\infty\frac{dyy^2}{e^y+1}+
              \frac{1}{2\pi ^2\xi _i}\int_0^\infty\frac{dyy^4}{e^y+1}
              \nonumber\\
           &=&\frac{\hat{n}_\nu}{T^3}\xi _i+
              \frac{1}{2\pi ^2\xi _i}\int_0^\infty\frac{dyy^4}{e^y+1}
              \nonumber\\
           &=&\frac{\hat{n}_\nu}{T^3}\xi _i
              +a\frac{\hat{n}_\nu}{T^3}\frac{1}{\xi _i},\label{F2}
 \end{eqnarray}
where
 \begin{eqnarray}
  \hat{n}_\nu\equiv\frac{T^3}{\pi ^2}\int_0^\infty
  \frac{dyy^2}{e^y+1},\hspace{5mm}
  a\equiv\frac{\int_0^\infty\frac{dyy^4}{e^y+1}}
  {2\int_0^\infty\frac{dyy^2}{e^y+1}}\simeq 6.47,
 \end{eqnarray}
and thus, we get
 \begin{eqnarray}
  \rho _\nu &=&\sum_{i=1}^3m_{\nu i}\hat{n}_\nu
               +\sum_{i=1}^3a\hat{n}_\nu\frac{T}{\xi _i}
               =\sum_{i=1}^3m_{\nu i}\hat{n}_\nu
               \left(1+\frac{a}{\xi _i^2}\right),\\
  \rho _{\mbox{{\scriptsize DE}}}&=&\rho _\nu+V(\phi _a)
  =\sum_{i=1}^3m_{\nu i}\hat{n}_\nu
   \left(1+\frac{a}{\xi _i^2}\right)+V(\phi _a),\\
  \frac{\partial\rho _{\mbox{{\scriptsize DE}}}}
       {\partial z}&=&\displaystyle\sum_{i=1}^3\left[\left(
                      \frac{\partial m_{\nu i}}{\partial z}
                      \hat{n}_\nu
                      +m_{\nu i}
                       \frac{\partial\hat{n}_\nu}{\partial z}
                       \right)\left(1+\frac{a}{\xi _i^2}\right)
                      -m_{\nu i}\hat{n}_\nu\frac{2a}{\xi _i^3}
                       \frac{\partial\xi _i}{\partial z}\right]
                      +\frac{\partial V(\phi _a)}{\partial z}.
  \label{DEz1}
 \end{eqnarray}
The stationary condition eq.(\ref{stationary}) leads to the relation:
 \begin{eqnarray}
  \frac{\partial V(\phi _a)}{\partial z}
  \frac{\partial z}{\partial\displaystyle\sum_{i=1}^3m_{\nu i}}
  &=&-\frac{\partial}{\partial\displaystyle\sum_{i=1}^3m_{\nu i}}
     \left[\sum_{j=1}^3m_{\nu j}\hat{n}_\nu
     \left(1+\frac{a}{\xi _j^2}\right)\right]\nonumber\\
  &=&-\hat{n}_\nu
     -\frac{\partial z}{\partial\displaystyle\sum_{i=1}^3m_{\nu i}}
      \frac{\partial}{\partial z}
      \left(\displaystyle\sum_{j=1}^3m_{\nu j}\hat{n}_\nu
            \frac{a}{\xi _j^2}\right)\nonumber\\
  &=&-\hat{n}_\nu
  -\frac{\partial z}{\partial\displaystyle\sum_{i=1}^3m_{\nu i}}
    \left[\displaystyle\sum_{j=1}^3\left(
    \frac{\partial m_{\nu j}}{\partial z}\hat{n}_\nu
           \frac{a}{\xi _j^2}
           +m_{\nu j}\frac{\partial\hat{n}_\nu}{\partial z}
            \frac{a}{\xi _j^2}
           -2m_{\nu j}\hat{n}_\nu\frac{a}{\xi _j^3}
            \frac{\partial\xi _j}{\partial z}\right)\right],
  \nonumber\\
 \end{eqnarray}
and thus, we have 
 \begin{eqnarray}
  \frac{\partial V(\phi _a)}{\partial z}&=&
  \displaystyle\sum_{i=1}^3\left[
  \frac{\partial m_{\nu i}}{\partial z}
  \hat{n}_\nu-a\left(
           \frac{\partial m_{\nu i}}{\partial z}\hat{n}_\nu
           \frac{1}{\xi _i^2}
           +m_{\nu i}\frac{\partial\hat{n}_\nu}{\partial z}
            \frac{1}{\xi _i^2}
           -2m_{\nu i}\hat{n}_\nu\frac{1}{\xi _i^3}
            \frac{\partial\xi _i}{\partial z}\right)\right].
  \label{Vz}
 \end{eqnarray}
Using eqs.(\ref{DEz1}) and (\ref{Vz}), the redshift dependence of the
dark energy is
 \begin{eqnarray}
  \frac{\partial\rho _{\mbox{{\scriptsize DE}}}}
       {\partial z}&=&\displaystyle\sum_{i=1}^3\left(
                      \frac{\partial m_{\nu i}}{\partial z}
                      \hat{n}_\nu
                      +m_{\nu i}
                       \frac{\partial\hat{n}_\nu}{\partial z}
                       \right)\left(1+\frac{a}{\xi _i^2}\right)
                      -\displaystyle\sum_{i=1}^3m_{\nu i}\hat{n}_\nu
                       \frac{2a}{\xi _i^3}
                       \frac{\partial\xi _i}{\partial z}
                       \nonumber\\
                   &&+\displaystyle\sum_{i=1}^3\left[
                      \frac{\partial m_{\nu i}}{\partial z}
                      \hat{n}_\nu-a\left(
                      \frac{\partial m_{\nu i}}{\partial z}\hat{n}_\nu
                      \frac{1}{\xi _i^2}
                      +m_{\nu i}\frac{\partial\hat{n}_\nu}{\partial z}
                      \frac{1}{\xi _i^2}
                     -2m_{\nu i}\hat{n}_\nu\frac{1}{\xi _i^3}
                      \frac{\partial\xi _i}{\partial z}\right)\right]
                      \nonumber\\
                   &=&\displaystyle\sum_{i=1}^3m_{\nu i}
                      \frac{\partial\hat{n}_\nu}{\partial z}.
                      \label{DEz}
 \end{eqnarray}

The equation of state parameter $w$ is 
 \begin{equation}
  w+1=\frac{[4-h(T)]\rho _\nu}{3\rho _{\mbox{{\scriptsize DE}}}},
 \end{equation}
where 
 \begin{equation}
  h(T)\equiv\frac{\displaystyle\sum_{i=1}^3\xi _i
            \frac{\partial F(\xi _i)}{\partial\xi _i}}
            {\displaystyle\sum_{j=1}^3F(\xi _j)}.
 \end{equation}
Using eq.(\ref{F2}),
 \begin{eqnarray}
  h(T)&=&\frac{\displaystyle\sum_{i=1}^3\xi _i
         \left(\frac{\hat{n}_\nu}{T^3}
         -a\frac{\hat{n}_\nu}{T^3}\frac{1}{\xi _i^2}\right)}
         {\displaystyle\sum_{j=1}^3\left(
         \frac{\hat{n}_\nu}{T^3}\xi _j
         +a\frac{\hat{n}_\nu}{T^3}\frac{1}{\xi _j}\right)}\nonumber\\
      &=&\frac{\displaystyle\sum_{i=1}^3\xi _i
               -\displaystyle\sum_{j=1}^3\frac{a}{\xi _j}}
         {\displaystyle\sum_{k=1}^3\xi _k
          +\displaystyle\sum_{l=1}^3\frac{a}{\xi _l}}\nonumber\\
      &=&\frac{1-\displaystyle\frac{1}
                                   {\displaystyle\sum_{i=1}^3\xi _i}
         \left(\displaystyle\sum_{j=1}^3\frac{a}{\xi _j}\right)}
         {1-\displaystyle\frac{1}{\displaystyle\sum_{k=1}^3\xi _k}
         \left(\displaystyle\sum_{l=1}^3\frac{a}{\xi _l}\right)}
         \nonumber\\
      &\simeq&1-\frac{2}{\displaystyle\sum_{i=1}^3\xi _i}
              \left(\displaystyle\sum_{j=1}^3\frac{a}{\xi _j}\right).
 \end{eqnarray}
Thus, we have
 \begin{eqnarray}
  w+1&=&\frac{\left[3+\displaystyle
              \frac{2}{\displaystyle\sum_{i=1}^3\xi _i}
              \left(\displaystyle\sum_{j=1}^3\frac{a}{\xi _j}\right)
              \right]\left(
              \displaystyle\sum_{k=1}^3m_{\nu k}\hat{n}_\nu
              +\displaystyle\sum_{l=1}^3a\hat{n}_\nu
              \frac{T}{\xi _l}\right)}
        {3\rho _{\mbox{{\scriptsize DE}}}}\nonumber\\
     &=&\frac{3\displaystyle\sum_{i=1}^3m_{\nu i}\hat{n}_\nu
              +3\displaystyle\sum_{i=1}^3a\hat{n}_\nu
              \frac{T}{\xi _i}
              +\frac{2}{\displaystyle\sum_{i=1}^3\xi _i}
              \left(\displaystyle\sum_{j=1}^3\frac{a}{\xi _j}\right)
              \displaystyle\sum_{k=1}^3m_{\nu k}\hat{n}_\nu}
              {3\rho _{\mbox{{\scriptsize DE}}}}\nonumber\\
     &&+\frac{\displaystyle\frac{2}{\displaystyle\sum_{i=1}^3\xi _i}
              \left(\displaystyle\sum_{j=1}^3\frac{a}{\xi _j}\right)
              \displaystyle\sum_{k=1}^3a\hat{n}_\nu
              \frac{T}{\xi _k}}
             {3\rho _{\mbox{{\scriptsize DE}}}},
 \end{eqnarray}
where the last term is negligible small because $\xi _i$ is much
 larger than $1$. Thus we get
 \begin{eqnarray}
  \rho _{\mbox{{\scriptsize DE}}}(w+1)
  &=&\frac{3\displaystyle\sum_{i=1}^3m_{\nu i}\hat{n}_\nu
     +3\displaystyle\sum_{i=1}^3a\hat{n}_\nu
     \frac{T}{\xi _i}
     +\frac{2T}{\displaystyle\sum_{i=1}^3m_{\nu i}}
     \left(\displaystyle\sum_{j=1}^3\frac{a}{\xi _j}\right)
     \displaystyle\sum_{k=1}^3m_{\nu k}\hat{n}_\nu}
     {3}\nonumber\\
  &=&\displaystyle\sum_{i=1}^3\left(m_{\nu i}\hat{n}_\nu
     +\frac{5a\hat{n}_\nu T}{3\xi _i}\right).\label{rhow}
 \end{eqnarray}
Differentiating eq.(\ref{rhow}) by the redshift parameter, we have
 \begin{equation}
  \frac{\partial\rho _{\mbox{{\scriptsize DE}}}}{\partial z}w
  +\frac{\partial\rho _{\mbox{{\scriptsize DE}}}}{\partial z}
  +\rho _{\mbox{{\scriptsize DE}}}\frac{\partial w}{\partial z}
  =\displaystyle\sum_{i=1}^3\left[
   \frac{\partial m_{\nu i}}{\partial z}\hat{n}_\nu
   +m_{\nu i}\frac{\partial\hat{n}_\nu}{\partial z}
   +\frac{5a}{3}\left(
    \frac{\partial\hat{n}_\nu}{\partial z}\frac{T}{\xi _i}
   +\frac{\partial T}{\partial z}\frac{\hat{n}_\nu}{\xi _i}
   -\frac{\hat{n}_\nu T}{\xi _i^2}
    \frac{\partial\xi _i}{\partial z}\right)\right],
 \end{equation}
which leads to
 \begin{eqnarray}
  \frac{\partial\rho _{\mbox{{\scriptsize DE}}}}{\partial z}w
  +\rho _{\mbox{{\scriptsize DE}}}\frac{\partial w}{\partial z}
  &=&\displaystyle\sum_{i=1}^3\left[
     \frac{\partial m_{\nu i}}{\partial z}\hat{n}_\nu
     +m_{\nu i}\frac{\partial\hat{n}_\nu}{\partial z}
     +\frac{5a}{3}\left(
      \frac{3\hat{n}_\nu}{z+1}\frac{T}{\xi _i}
     +T_0\frac{\hat{n}_\nu}{\xi _i}
     -\frac{\hat{n}_\nu T}{\xi _i^2}
      \frac{\partial\xi _i}{\partial z}\right)\right]
     -\frac{\partial\rho _{\mbox{{\scriptsize DE}}}}{\partial z}\nonumber\\  
  &=&\displaystyle\sum_{i=1}^3\left[
     \frac{\partial m_{\nu i}}{\partial z}\hat{n}_\nu
     +m_{\nu i}\frac{\partial\hat{n}_\nu}{\partial z}
     +\frac{5a\hat{n}_\nu}{3}\left(
      \frac{4T_0}{\xi _i}
     -\frac{T}{\xi _i^2}
      \frac{\partial\xi _i}{\partial z}\right)\right]
     -\frac{\partial\rho _{\mbox{{\scriptsize DE}}}}{\partial z}.
 \end{eqnarray}
Then, using the relation (\ref{DEz}), we have
 \begin{equation}
  \frac{\partial\rho _{\mbox{{\scriptsize DE}}}}{\partial z}w
  +\rho _{\mbox{{\scriptsize DE}}}\frac{\partial w}{\partial z}
  =\displaystyle\sum_{i=1}^3\left[
     \frac{\partial m_{\nu i}}{\partial z}\hat{n}_\nu
     +m_{\nu i}\frac{\partial\hat{n}_\nu}{\partial z}
     +\frac{5a\hat{n}_\nu}{3}\left(
     \frac{4T_0}{\xi _i}
    -\frac{T}{\xi _i^2}
     \frac{\partial\xi _i}{\partial z}\right)\right]
    -\displaystyle\sum_{i=1}^3m_{\nu i}
     \frac{\partial\hat{n}_\nu}{\partial z}.
  \label{numerator}
 \end{equation}
Since the speed of sound squared in the dark energy is given by
 \begin{eqnarray}
  c_s^2&\equiv&\frac{\partial p}
                    {\partial\rho _{\mbox{{\scriptsize DE}}}}\nonumber\\
       &=&\frac{\partial (w\rho _{\mbox{{\scriptsize DE}}})}
               {\partial\rho _{\mbox{{\scriptsize DE}}}}\nonumber\\
       &=&\frac{
          \frac{\partial\rho _{\mbox{{\scriptsize DE}}}}{\partial z}w
                +\rho _{\mbox{{\scriptsize DE}}}
                 \frac{\partial w}{\partial z}}
          {\frac{\partial\rho _{\mbox{{\scriptsize DE}}}}
                {\partial z}},
 \end{eqnarray}
using eqs.(\ref{DEz}) and (\ref{numerator}), we get finally
 \begin{eqnarray}
  c_s^2&=&\frac{\displaystyle\sum_{i=1}^3\left[
                \frac{\partial m_{\nu i}}{\partial z}\hat{n}_\nu
                +m_{\nu i}\frac{\partial\hat{n}_\nu}{\partial z} 
                +\frac{5a\hat{n}_\nu}{3}\left(
                 \frac{4T_0}{\xi _i}
                -\frac{T}{\xi _i^2}
                 \frac{\partial\xi _i}{\partial z}\right)\right]
                -\displaystyle\sum_{i=1}^3m_{\nu i}
                        \frac{\partial\hat{n}_\nu}{\partial z}}
               {\displaystyle\sum_{i=1}^3m_{\nu i}
                        \frac{\partial\hat{n}_\nu}{\partial z}}
                        \nonumber\\
       &=&\frac{\displaystyle\sum_{i=1}^3
          \frac{\partial m_{\nu i}}{\partial z}\hat{n}_\nu}
          {\displaystyle\sum_{i=1}^3m_{\nu i}
          \frac{\partial \hat{n}_\nu }{\partial z}}
          +\frac{\frac{5}{3}a\hat{n}_\nu\displaystyle\sum_{i=1}^3
           \left(\frac{5T_0}{\xi _i}
          -\frac{T}{\xi _i^2}
          \frac{\partial m_{\nu i}}{\partial z}\right)}
          {\displaystyle\sum_{i=1}^3m_{\nu i}
          \frac{\partial\hat{n}_\nu }{\partial z}}.
 \end{eqnarray}
It is easy to see that if the following relation is satisfied, the
 speed of sound squared becomes positive:
 \begin{equation}
  \displaystyle\sum_{i=1}^3
  \frac{\partial m_{\nu i}}{\partial z}
  \left(1-\frac{5aT^2}{3m_{\nu i}^2}\right)
  +\frac{25aT_0^2(z+1)}{3}\displaystyle\sum_{i=1}^3
  \frac{1}{m_{\nu i}}>0.
 \end{equation}

\end{document}